\documentclass[aps,pra,twocolumn,showpacs]{revtex4}
\usepackage{amsfonts}
\usepackage{amssymb}
\usepackage{amsmath}
\usepackage{graphicx}
\usepackage{txfonts}

\begin{document}

\title{Nondestructive identification of the Bell diagonal state}
\author{Jia-sen Jin}
\author{Chang-shui Yu}
\email{quaninformation@sina.com;ycs@dlut.edu.cn}
\author{He-shan Song}
\email{hssong@dlut.edu.cn}
\affiliation{School of Physics and Optoelectronic Technology,\\
Dalian University of Technology, Dalian 116024 China }
\date{\today}

\begin{abstract}
We propose a scheme for identifying an unknown Bell diagonal state.
In our scheme the measurements are performed on the probe qubits
instead of the Bell diagonal state. The distinguished advantage is
that the quantum state of the evolved Bell diagonal state ensemble
plus probe states will still collapse on the original Bell diagonal
state ensemble after the measurement on probe states, i.e. our
identification is quantum-state nondestructive. It is also shown
finally how to realize our scheme in the framework of cavity
electrodynamics.
\end{abstract}

 \maketitle

\section{Introduction}

Entanglement is not only an essential feature of quantum mechanics,
which distinguishes the quantum from classical world, but also a
great resource in the fields of quantum information and quantum
computation [1-3]. Particularly, entangled qubits prepared in the
pure maximally entangled states, \emph{i.e}., the Bell states, are
required by many quantum information processes [4,5]. However, in
the real world, a pure state in a quantum system will always evolves
to a mixed one due to the unavoidable interactions with the
environment. Thus, for practical purpose, applications of quantum
information processes utilizing the mixed states are under
consideration.

Among the bipartite entangled mixed states, the Bell diagonal state
(BDS) plays an important role in quantum information processing. It
is widely used in the processes of quantum teleportation \cite{m1},
quantum entanglement purification [7,8], quantum key distribution
\cite{m4}, etc. Moreover, the BDS is a simple but significant
example in studying the nonclassical correlation of a quantum mixed
state [10-12], since there always exists a local transformation
which can transform the given mixed state to a corresponding Bell
diagonal form \cite{Cen}. Therefore, identification of an unknown
BDS is of great importance. Conventionally, the identification is
achieved by the so-called state tomography technique [14-16], which
performs the projection measurements on the unknown state directly,
and repeats the measurements on many copies of state. It is a
drawback that after the projection measurements the state to be
measured will collapse to one of the measurement basis. Thus the
original state will be destroyed and become useless.

Recently, schemes for quantum non-demolition measurement are
proposed to detect unknown quantum state [17-19], by which the
detected state will not be destroyed after the measurement. In this
paper we present an alternative scheme for identifying an unknown
BDS. In our scheme, we do not perform the projective measurements on
the BDS directly, but on the probe qubits. According to the
measurement outcomes of the probe qubits, we can acquire all the
information of the unknown BDS. The distinguished advantage of our
scheme is that the BDS is not destroyed by the measurements, since
the evolved BDS plus the probe states will collapse to the original
BDS after the measurement on the probe qubits. Contrast to the
identification scheme with state tomography technique, which is
achieved by sacrificing numerous copies of the unknown state, our
scheme is economic and the resulting BDS is recyclable. The paper is
organized as follows. In Sec. II, we explicitly demonstrate our
scheme in theory. In Sec. III, we discuss the experimental
realization of our scheme in the framework of cavity quantum
electrodynamics (QED). The conclusion is drawn finally.

\section{Scheme for identification of unknown Bell diagonal state}

In this section, we will illustrate our scheme explicitly. The BDS
is a mixture of the well-known Bell states, it is parameterized by
four real numbers $c_1,c_2,c_3,c_4\in[0,1]$, which satisfy the
normalizing condition $\sum_ic_i=1$. Generally, a BDS can be
described as $\rho_{12}=\sum_ic_i|\Psi_i\rangle\langle\Psi_i|$,
where
$|\Psi_{1}\rangle=(|1\rangle_1|0\rangle_2+|0\rangle_1|1\rangle_2)/\sqrt{2}$,
$|\Psi_{2}\rangle=(|1\rangle_1|0\rangle_2-|0\rangle_1|1\rangle_2)/\sqrt{2}$,
$|\Psi_{3}\rangle=(|1\rangle_1|1\rangle_2+|0\rangle_1|0\rangle_2)/\sqrt{2}$,
and
$|\Psi_{4}\rangle=(|1\rangle_1|1\rangle_2-|0\rangle_1|0\rangle_2)/\sqrt{2}$
are the Bell states, the subscript outside the ket denotes the label
of qubit. Here $|0\rangle=[1,0]^\mathrm{T}$ and
$|1\rangle=[0,1]^\mathrm{T}$ are the computational basis, the
superscript $\mathrm{T}$ denotes the matrix transpose.

In order to identify an unknown BDS, we need a probe qubit (labeled
3). The probe qubit interacts with the BDS and extracts the
information from the BDS. We accomplish our scheme by three steps,
in each step we can build an equation between the unknown parameters
and the observable of the probe qubit. With the three steps done, we
will obtain three independent linear equations, from which we can
calculate the parameters. The three steps of our scheme are
demonstrated in the following paragraphs.

\emph{Step 1.} Assume that the probe qubit is in state
$|0\rangle_3$, thus the initial state of the joint system consists
of the BDS and the probe qubit is given by
$\rho^0=\rho_{12}\otimes|0\rangle_3\langle0|$. We perform an unitary
operation $U^1$, given as follows, on the joint three-qubit state,
\begin{equation}
U^1=\frac{1}{2}\left(
                 \begin{array}{rrrrrrrr}
                   1 & 0 & 0 & -i & 0 & -i & -1 & 0 \\
                   0 & 1 & -i & 0 & -i & 0 & 0 & -1 \\
                   0 & -i & 1 & 0 & -1 & 0 & 0 & -i \\
                   -i & 0 & 0 & 1 & 0 & -1 & -i & 0 \\
                   0 & -i & -1 & 0 & 1 & 0 & 0 & -i \\
                   -i & 0 & 0 & -1 & 0 & 1 & -i & 0 \\
                   -1 & 0 & 0 & -i & 0 & -i & 1 & 0 \\
                   0 & -1 & -i & 0 & -i & 0 & 0 & 1 \\
                 \end{array}
               \right)
               \label{as}
\end{equation}
As a result the state of the joint system evolves to
$\rho^1=U^1\rho^0U^{1\dagger}$. We can obtain the reduced density
matrix of the probe qubit by tracing over qubits 1 and 2 as follows,
\begin{equation}
\rho_3^1=\left(
           \begin{array}{cc}
             c_1+c_3 & 0 \\
             0 & c_2+c_4 \\
           \end{array}
         \right).
\end{equation}
One can find that the information of the BDS is carried by the probe
qubit. Performing a $\sigma^z$ measurement on the probe qubit, we
can obtain the following equation between the unknown parameters and
the observable of the probe qubit,
\begin{equation}
M^1=\mathrm{Tr}(\sigma^z_3\rho^1_3)=c_1+c_3-c_2-c_4.
\end{equation}
Tracing over the probe qubit we can obtain the reduced density
matrix of the resulting BDS as follows,
\begin{equation}
\rho^1_{12}=\frac{1}{2}\left(
                     \begin{array}{cccc}
                       c_1+c_4 & 0 & 0 & c_1-c_4 \\
                       0 & c_3+c_2 & c_3-c_2 & 0 \\
                       0 & c_3-c_2 & c_3+c_2 & 0 \\
                       c_1-c_4 & 0 & 0 & c_1+c_4 \\
                     \end{array}
                   \right).
\end{equation}
Note that underwent the $U^1$ operation, the resulting BDS becomes
different from the original state because the $|\Psi_1\rangle$ and
$|\Psi_3\rangle$ ingredients have exchanged mutually. Fortunately,
we can recover it to the original form by repeating the
above-mentioned process once more with a new probe qubit to exchange
$|\Psi_1\rangle$ and $|\Psi_3\rangle$ ingredients again. It is
interesting that at the end of the recovering process, we can obtain
the same reduced density matrix of the new probe qubit as shown in
Eq. (2) and consequently yield equation Eq. (3) from the new
resulting probe qubit.

\emph{Step 2}. In this step, the probe qubit is also initialized in
$|0\rangle_3$, thus the joint system is in state
$\rho_{12}\otimes|0\rangle_3\langle 0|$. We perform an unitary
operation named $U^2$ on the joint three-qubit state, $U^2$ has the
following form,
\begin{equation}
U^2=\frac{1}{2}\left(
                 \begin{array}{rrrrrrrr}
                   1 & 0 & 0 & i & 0 & i & 1 & 0 \\
                   0 & 1 & -i & 0 & -i & 0 & 0 & 1 \\
                   0 & -i & 1 & 0 & -1 & 0 & 0 & i \\
                   i & 0 & 0 & 1 & 0 & -1 & -i & 0 \\
                   0 & -i & -1 & 0 & 1 & 0 & 0 & i \\
                   i & 0 & 0 & -1 & 0 & 1 & -i & 0 \\
                   1 & 0 & 0 & -i & 0 & -i & 1 & 0 \\
                   0 & 1 & i & 0 & i & 0 & 0 & 1 \\
                 \end{array}
               \right).
\end{equation}
After $U^2$ operation, the probe qubit evolves to the following
form,
\begin{equation}
\rho^2_3=\left(
           \begin{array}{cc}
             c_1+c_4 & 0 \\
             0 & c_2+c_3 \\
           \end{array}
         \right).
\end{equation}
Performing a $\sigma^z$ measurement on the probe qubit, we can
obtain the following equation,
\begin{equation}
M^2=\mathrm{Tr}(\sigma^z_3\rho_3^2)=c_1+c_4-c_2-c_3.
\end{equation}

The resulting BDS underwent the $U^2$ operation is given as follows,
\begin{equation}
\rho^2_{12}=\frac{1}{2}\left(
                         \begin{array}{cccc}
                           c_3+c_1 & 0 & 0 & c_3-c_1 \\
                           0 & c_4+c_2 & c_4-c_2 & 0 \\
                           0 & c_4-c_2 & c_4+c_2 & 0 \\
                           c_3-c_1 & 0 & 0 & c_3+c_1 \\
                         \end{array}
                       \right),
\end{equation}
Similar to step 1, we can transform the resulting BDS to the
original form by performing $U^2$ on the joint system which is
composed of the resulting BDS and a new probe qubit. Again the new
resulting probe qubit ensemble will carry the information of the
unknown parameters.

\emph{Step 3.} We perform an unitary operation $U^3$ on the joint
BDS and probe qubit system, $U^3$ is given as follows,
\begin{equation}
U^3=\left(
      \begin{array}{cccccccc}
        -i & 0 & 0 & 0 & 0 & 0 & 0 & 0 \\
        0 & i & 0 & 0 & 0 & 0 & 0 & 0 \\
        0 & 0 & 1 & 0 & 0 & 0 & 0 & 0 \\
        0 & 0 & 0 & 1 & 0 & 0 & 0 & 0 \\
        0 & 0 & 0 & 0 & 1 & 0 & 0 & 0 \\
        0 & 0 & 0 & 0 & 0 & 1 & 0 & 0 \\
        0 & 0 & 0 & 0 & 0 & 0 & i & 0 \\
        0 & 0 & 0 & 0 & 0 & 0 & 0 & -i \\
      \end{array}
    \right).
\end{equation}
Different from the previous two steps, here the probe qubit is
initialized in the superposition state
$(|0\rangle_3+|1\rangle_3)/\sqrt{2}$. After $U^3$ operation, the
probe qubit evolves to the following form,
\begin{equation}
\rho^3_3=\frac{1}{2}\left(
                      \begin{array}{cc}
                        1 & c_1+c_2-c_3-c_4 \\
                        c_1+c_2-c_3-c_4 & 1 \\
                      \end{array}
                    \right).
\end{equation}
Now, performing a $\sigma^x$ measurement on the probe qubit, we can
obtain the following equation,
\begin{equation}
M^3=\mathrm{Tr}(\sigma^x_3\rho_3^3)=c_1+c_2-c_3-c_4.
\end{equation}
In the meantime, the resulting BDS has the following form,
\begin{equation}
\rho^3_{12}=\frac{1}{2}\left(
                         \begin{array}{cccc}
                           c_3+c_4 & 0 & 0 & c_4-c_3 \\
                           0 & c_1+c_2 & c_1-c_2 & 0 \\
                           0 & c_1-c_2 & c_1+c_2 & 0 \\
                           c_4-c_3 & 0 & 0 & c_3+c_4 \\
                         \end{array}
                       \right).
\end{equation}
To recover this BDS back to the original state, we repeat $U^3$ on
the joint system composed of the resulting BDS and a new probe qubit
ensemble in state $(|0\rangle_3+|1\rangle_3)/\sqrt{2}$. Obviously,
the new resulting probe qubit ensemble will also carry the
information of the unknown BDS.

Combining equations (3), (7), and (11), and taking into account the
normalizing condition $\sum_ic_i=1$, we can work out the parameters
as:
\begin{equation}
c_1=\frac{M^1+M^2+M^3+1}{4},
\end{equation}
\begin{equation}
c_2=\frac{1-M^1-M^2+M^3}{4},
\end{equation}
\begin{equation}
c_3=\frac{1+M^1-M^2-M^3}{4},
\end{equation}
\begin{equation}
c_4=\frac{1-M^1+M^2-M^3}{4}.
\end{equation}

Now we have succeeded in identifying an unknown BDS with the help of
the probe qubit ensembles. Notably, due to the recovering process in
each step, the final BDS is the same as the initial state.

It is necessary give a discussion on the principles of our scheme.
We emphasize that our scheme is based on the ensemble viewpoint, by
which the BDS can be considered as a mixture of the four Bell
states. The mixing proportion of each Bell state is denoted by the
parameter $c_i$. Each pair of qubits 1 and 2 fetching from the BDS
ensemble will be randomly in one of the four Bell states. Without
loss of generality, we take step 1 as an example to show how the
probe qubit can extract information from the BDS ensemble. The
expression of $U^1$ can be rewritten as
$U^1=|\psi_{2,0}\rangle\langle\psi_{2,0}|+|\psi_{2,1}\rangle\langle\psi_{2,1}|+|\psi_{4,0}\rangle\langle\psi_{4,0}|+|\psi_{4,1}\rangle\langle\psi_{4,1}|-i|\psi_{1,0}\rangle\langle\psi_{3,1}|-i|\psi_{1,1}\rangle\langle\psi_{3,0}|-i|\psi_{3,0}\rangle\langle\psi_{1,1}|-i|\psi_{3,1}\rangle\langle\psi_{1,0}|
$, where $|\psi_{i,j}\rangle=|\Psi_i\rangle\otimes|j\rangle_3$. One
can find that if the state of qubits 1 and 2 is $|\Psi_2\rangle$ or
$|\Psi_4\rangle$, it will remain unchanged and the probe qubit 3
will stay in $|0\rangle_3$; if the state of qubits 1 and 2 is
$|\Psi_1\rangle$ ($|\Psi_3\rangle$), it will change to
$|\Psi_3\rangle$ ($|\Psi_1\rangle$) and flip the probe qubit state
from $|0\rangle_3$ to $|1\rangle_3$. Repeatedly perform $U^1$ on the
joint three-qubit state by fetching new qubits from the BDS and the
probe qubit ensemble, the resulting probe qubit ensemble will end in
a mixed state ensemble which reveals the information of $c_1$ and
$c_3$ through the appearance probability of $|1\rangle_3$. To
transform the resulting BDS back to the original form, we only need
to repeat this process once more to make a simple exchange of
$|\Psi_1\rangle$ and $|\Psi_3\rangle$. In steps 2 and 3, our scheme
works similarly. As a consequence, the information of the unknown
BDS is transferred to the probe ensembles. It is interesting that
the resulting probe ensembles produced by the recover process are
also useful.

Let us look back to the expressions of $U^1$, $U^2$, and $U^3$.
These operators are  essentially tripartite manipulations on qubits,
and they can be formally factorized as $U^i=(U^i_{13}\otimes
I_2)(I_1\otimes U^i_{23}) (i=1,2,3)$, where $I_1$ and $I_2$ are the
identity operators of subsystems 1 and 2, respectively. The
bipartite operations $U^i_{13}$ and $U^i_{23}$ are given as follows,
\begin{equation}
U^{1}_{13}=U^1_{23}=\frac{1}{\sqrt{2}}\left(
                                      \begin{array}{cccc}
                                        1 & 0 & 0 & -i \\
                                        0 & 1 & -i & 0 \\
                                        0 & -i & 1 & 0 \\
                                        -i & 0 & 0 & 1 \\
                                      \end{array}
                                    \right),
\end{equation}
\begin{equation}
U^2_{13}=U^2_{23}=\frac{1}{\sqrt{2}}\left(
                                      \begin{array}{cccc}
                                        1 & 0 & 0 & i \\
                                        0 & 1 & -i & 0 \\
                                        0 & -i & 1 & 0 \\
                                        i & 0 & 0 & 1 \\
                                      \end{array}
                                    \right),
\end{equation}
\begin{equation}
U^3_{13}=U^3_{23}=\frac{1}{\sqrt{2}}\left(
                                      \begin{array}{cccc}
                                        1-i & 0 & 0 & 0 \\
                                        0 & 1+i & 0 & 0 \\
                                        0 & 0 & 1+i & 0 \\
                                        0 & 0 & 0 & 1-i \\
                                      \end{array}
                                    \right).
\end{equation}

Based on these factorizations we can accomplish each step by
sequentially performing bipartite manipulation $U^i_{13}$ on qubits
1 and 3, and $U^i_{23}$ on qubits 2 and 3. That is to say we can
perform only bipartite manipulations in the whole processing of our
scheme, instead of tripartite manipulations which is difficult to
realize in experiments. The procedures are given as follows. Suppose
that qubit 1 together with qubit 3 locates at place A, and qubit 3
locates at place B. In each step, we first perform operations
$U^{i}_{13}$ on qubits 1 and 3, next send qubit 3 to place B, and
then perform operations $U^i_{23}$ on qubits 2 and 3. Finally, we
perform measurements on the probe qubit.

\section{Identification of BDS in experimental scenario}

\begin{figure}[tbp]
\includegraphics[width=1\columnwidth]{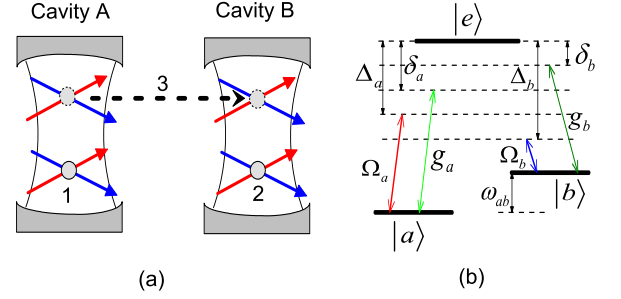}
\caption{(Color online) (a) Schematic illustration for
identification of unknown BDS based on the cavity QED system. The
probe atom 3 first interacts with atom 1 in cavity A and then is
sent to cavity B to interacts with atom 2. Each atom in the cavity
is driven by two classical lasers. (b) Atomic levels and
transitions.} \label{1}
\end{figure}

In this section we will discuss experimental realization of our
scheme in the framework of cavity QED system. This experimental
scenario is based on the case that two qubits are separated into
different places, since two-qubit manipulation is more feasible than
the three-qubit manipulation. The schematic illustration is shown in
Fig. \ref{1}(a). Assume that the BDS ensemble is shared by two
participators, each of whom has an optical cavity A and B,
respectively. The particles in the ensemble are considered to be
three-level atoms with two ground states $|a\rangle$ and $|b\rangle$
and an excited state $|e\rangle$, see Fig. \ref{1}(b). The
long-lived levels $|a\rangle$ and $|b\rangle$ represent states
$|0\rangle$ and $|1\rangle$, respectively. The probe atoms are
identical to those in the BDS ensemble. The cavity couples the
atomic transitions $|a\rangle\leftrightarrow|e\rangle$ and
$|b\rangle\leftrightarrow|e\rangle$ with the coupling strength $g_a$
and $g_b$, respectively. Additionally, two external driving lasers
couple the transitions $|a\rangle\leftrightarrow|e\rangle$ and
$|b\rangle\leftrightarrow|e\rangle$ with the Rabi frequencies
$\Omega_a$ and $\Omega_b$, respectively. All the atoms couple to the
cavities via the same mechanism.

To start the scheme, we pick up a pair of entangled atoms from the
BDS ensemble and put them into the corresponding cavities. The probe
atom 3 is sent into cavity A firstly. The Hamiltonian in cavity A
can be written as follows,
\begin{eqnarray}
H_A&=&\sum_{j=1,3}\omega_e|e\rangle_j\langle
e|+\omega_{ab}|b\rangle_j\langle
b|+\omega_ca^{\dagger}a\cr\cr&&+(\Omega_ae^{-i\omega_at}+g_aa)|e\rangle_j\langle
a|\cr\cr&&+(\Omega_be^{-i\omega_bt}+g_ba)|e\rangle_j\langle b|,
\end{eqnarray}
where $\omega_e$ is the energy of $|e\rangle$ while $\omega_{ab}$ is
the energy of $|b\rangle$, $\omega_a$ ($\omega_b$) is the frequency
of the driver laser with Rabi frequency $\Omega_a$ ($\Omega_b$), and
$a$ is the annihilation operator of the cavity.

By setting $\delta_1=\omega_{ab}-(\omega_a-\omega_b)/2$, we switch
to an interaction picture with respect to
$H_0=\sum_{j=1,3}\omega_e|e\rangle_j\langle
e|+(\omega_{ab}-\delta_1)|b\rangle_j\langle
b|+\omega_ca^{\dagger}a$. Under the large detuning condition
$|\delta_a|, |\delta_b|, |\Delta_a|, |\Delta_b|\gg|g_a|, |g_b|,
|\Omega_a|, |\Omega_b|$, where $\delta_a=\omega_e-\omega_c$,
$\delta_b=\omega_e-\omega_c-\omega_{ab}+\delta_1$,
$\Delta_a=\omega_e-\omega_a$,
$\Delta_b=\omega_e-\omega_b-\omega_{ab}+\delta_1$, we can
adiabatically eliminated the excited state $|e\rangle_j$ [20-23]. If
there are no photons in the cavity and the detunings satisfy
$|\delta_a-\Delta_b+g_a^2/\delta_a|,
|\delta_b-\Delta_a+g_a^2/\delta_a|\gg|\Omega_ag_b/\Delta_a|,
|\Omega_bg_a/\Delta_b|$ (we have assumed
$g_a^2/\delta_a=g_b^2/\delta_b$), considering the subspace without
real photons, we deduce the effective Hamiltonian as
\begin{equation}
H_{\mathrm{eff}}=\frac{2(J_1\sigma_1^++J_2\sigma_1^-)(J_1\sigma_3^-+J_2\sigma_3^+)}{\delta_a-\Delta_b+g_a^2/\delta_a}+B(\sigma_1^z+\sigma_3^z),
\end{equation}
where $\sigma_i^+=|b\rangle_i\langle a|$ and
$\sigma_i^-=|a\rangle_i\langle b|$, the coefficients are given as
\begin{equation}
J_1=\frac{g_a\Omega_b}{2}(\frac{1}{\delta_a}+\frac{1}{\Delta_b}),
\end{equation}
\begin{equation}
J_2=\frac{g_b\Omega_a}{2}(\frac{1}{\delta_b}+\frac{1}{\Delta_a}),
\end{equation}
\begin{eqnarray}
B&=&\frac{\Omega_a^2\Omega_b^2/4}{\delta_a-\delta_b}(\frac{1}{\Delta_a}+\frac{1}{\Delta_b})^2+\frac{\Omega_b^2g_b^2/4}{\delta_b-\Delta_a+g_a^2/\delta_a}(\frac{1}{\Delta_b}+\frac{1}{\delta_b})^2
\cr\cr&&-\frac{\Omega_a^2g_a^2/4}{\delta_a-\Delta_a+g_a^2/\delta_a}(\frac{1}{\Delta_a}+\frac{1}{\delta_a})^2
\cr\cr&&+\frac{1}{2}(\frac{\Omega_b^2}{\Delta_b}-\frac{\Omega_a^2}{\Delta_a}+\delta_1).
\end{eqnarray}
The effective magnetic field $B$ can be tuned to be very close to
zero by varying $\delta_1$. For
$\Omega_b=g_b\Omega_a(\frac{1}{\Delta_a}+\frac{1}{\delta_b})/[g_a(\frac{1}{\delta_a}+\frac{1}{\Delta_b})]$,
we can get the final effective Hamiltonian as follows,
\begin{equation}
H_{xx}=\lambda_x\sigma_1^x\sigma_3^x.
\end{equation}
where $\lambda_x=2J_1^2/(\delta_a-\Delta_b+g_a^2/\delta_a)$. It is
obvious to see that the unitary time-evolution operator
$e^{-iH_{xx}t}$ is in accordance with the unitary operator
$U_{13}^1$ at $t=\frac{(2n+1)\pi}{4\lambda_x},n=0,1,2...$, thus we
realize the unitary operation $U_{13}$. To realize the operation
$U_{23}^2$, we send the probe atom to cavity B, and drive atoms 2
and 3 with the same lasers as done in cavity A.

If we select the Rabi frequency as
$\Omega_b=-g_b\Omega_a(\frac{1}{\Delta_a}+\frac{1}{\delta_b})/[g_a(\frac{1}{\delta_a}+\frac{1}{\Delta_b})]$,
we can obtain
\begin{equation}
H_{yy}=\lambda_y\sigma_1^y\sigma_3^y.
\end{equation}
where $\lambda_y=2J_1^2/(\delta_a-\Delta_b+g_a^2/\delta_a)$. At time
$t=\frac{(2n+1)\pi}{4\lambda_y},n=0,1,2...$, the time-evolution
unitary operator $e^{-iH_{yy}t}$ coincides with the operator
$U^2_{13}$. Then sent the probe atom into cavity B, we can realize
the unitary operator $U^2_{23}$ by controlling the interaction time.

In order to realize the operations $U^3_{13}$ and $U^3_{23}$, we
choose the laser frequencies as $\omega_a=\omega_b=\omega$. We
switch to an interaction picture with respect to
$H_0=\sum_{j=1,3}\omega_e|e\rangle_j\langle
e|+(\omega_{ab}-\tilde{\delta}_1)|b\rangle_j\langle
b|+\omega_ca^{\dagger}a$, where the detuning $\tilde{\delta}_1$ is
introduced to tune the effective magnetic field. Under the large
detuning condition we can adiabatically eliminated the excited
states. If there are no photons in the cavity and the detunings
satisfy $|\delta_a-\Delta_a+g_a^2/\delta_a|,
|\delta_b-\Delta_b+g_a^2/\delta_a|\gg|\Omega_ag_a/\Delta_a|,
|\Omega_bg_b/\Delta_b|$ (assuming $g_a^2/\delta_a=g_b^2/\delta_b$),
considering a subspace with no real photons we can obtain the
following effective Hamiltonian,
\begin{eqnarray}
H_{\mathrm{eff}}&=&\frac{2(\tilde{J}_1|a\rangle_1\langle
a|+\tilde{J}_2|b\rangle_1\langle b|)(\tilde{J}_1|a\rangle_2\langle
a|+\tilde{J}_2|b\rangle_2\langle
b|)}{\delta_a-\Delta_a+g^2_a/\delta_a}\cr\cr&&+\tilde{B}(\sigma_1^z+\sigma^z_3),
\end{eqnarray}
where $\sigma^z_i=|b\rangle_i\langle b|-|a\rangle_i\langle a|$, and
the coefficients are given as follows,
\begin{equation}
\tilde{J}_1=\frac{\Omega_ag_a}{2}(\frac{1}{\Delta_a}+\frac{1}{\delta_a}),
\end{equation}
\begin{equation}
\tilde{J}_2=\frac{\Omega_bg_b}{2}(\frac{1}{\Delta_b}+\frac{1}{\delta_b}),
\end{equation}
\begin{eqnarray}
\tilde{B}&=&\frac{1}{2}(\frac{\Omega^2_b}{\Delta_b}+\tilde{\delta}_1-\frac{\Omega^2_a}{\Delta_a})+\frac{1}{2}[\frac{\Omega_b^2g_a^2/4}{\delta_a-\Delta_b+g_a^2/\delta_a}(\frac{1}{\delta_a}+\frac{1}{\Delta_b})^2\cr\cr&&-\frac{\Omega_a^2g_b^2/4}{\delta_b-\Delta_a+g_a^2/\delta_a}(\frac{1}{\delta_b}+\frac{1}{\Delta_a})^2]\cr\cr&&+\frac{\Omega_a^2\Omega_b^2/4}{\Delta_a-\Delta_b}(\frac{1}{\Delta_a}+\frac{1}{\Delta_b})^2.
\end{eqnarray}
The effective magnetic field can be tuned to zero by varying
$\tilde{\delta}_1$. For
$\Omega_b=-\Omega_ag_a(\frac{1}{\Delta_a}+\frac{1}{\delta_a})/[g_b(\frac{1}{\Delta_b}+\frac{1}{\delta_b})]$,
we can obtain the following effective Hamiltonian,
\begin{equation}
H_{zz}=\lambda_z\sigma_1^z\sigma_3^z,
\end{equation}
where $\lambda_z=2\tilde{J}_1/(\delta_a-\Delta_a+g_a^2/\delta_a)$.
It is obvious to see that the time-evolution unitary operator
$e^{-iH_{zz}t}$ coincides with $U_{13}^3$ at time points
$t=\frac{(2n+1)\pi}{4\lambda_z}$ $(n=0,1,2,...)$. Thus we have
realized the operation $U^3_{13}$ in cavity A, in the same way we
can realize the operation $U^3_{23}$ in cavity B by sending the
probe atom into cavity B.

To confirm the validity of our approximation, we numerically
simulate the dynamics generated by the full Hamiltonian and compare
it to the the dynamics generated by the effective Hamiltonian. In
Fig. 2(a), we have plotted the time evolution of
$\langle\sigma_3^z\rangle$. The numerical results show that the
performance of $\langle\sigma_3^z\rangle$ under the full Hamiltonian
and that under $H_{xx}$ agree with each other reasonably well.
Similar agreement can also be seen in Fig. 2(b) and (c). Therefore,
our effective model is valid.

\begin{figure}[tbp]
\includegraphics[width=1\columnwidth]{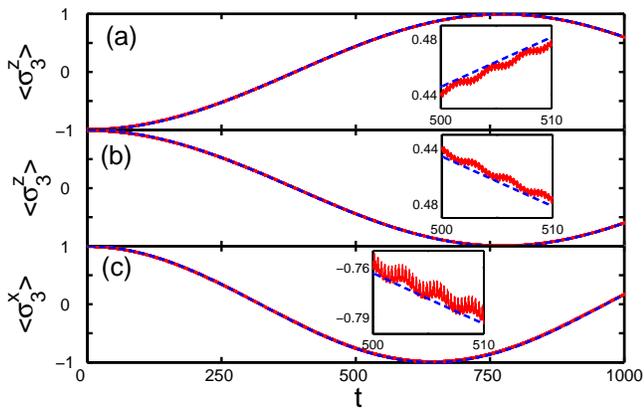}
\caption{(Color online) Time evolutions of atom 3 calculated using
full Hamiltonian (dashed blue line) and effective Hamiltonians
(solid red line). The effective Hamiltonians for (a), (b), and (c)
are $H_{xx}$, $H_{yy}$, and $H_{zz}$, respectively. The initial
states are chosen as $0.5(|a\rangle_1\langle a|+|b\rangle_1\langle
b|)\otimes|a\rangle_3\langle a|$ for (a) and (b), and
$0.25(|a\rangle_1\langle a|+|b\rangle_1\langle
b|)\otimes(|a\rangle+|b\rangle)_3(\langle a|+\langle b|)$ for (c).
The parameters are chosen as $g_a=1$, $\Omega_a=5g_a$,
$\delta_a=102g_a$, $\delta_b=122g_a$, $\Delta_a=120g_a$, and
$\Delta_b=100g_a$ for (a) and (b); $g_a=1$, $\Omega_a=5g_a$,
$\delta_a=102g_a$, $\delta_b=122g_a$, $\Delta_a=100g_a$, and
$\Delta_b=120g_a$ for (c). The inset shows the behaviors of the two
curves in detail with time $t\in[500,510]$ (in units of $1/g_a$).}
\label{2}
\end{figure}

So far, we have realized all the unitary operations described by
Eqs. (17)-(19) in the cavity QED system. Since the qubits are
encoded in the ground atomic states and there is no real photons in
the cavity, this experimental scenario is robust against the
dissipative effects.

Here we give a brief discussion on the experimental feasibility of
the presented scheme. For an experimental implementation, the
effective coupling strengths $\lambda_x$, $\lambda_y$, and
$\lambda_z$ should be much larger than the cavity leaky rate
$\Gamma_C$ and the atomic spontaneous rate $\Gamma_E$. This
requirements can be satisfied in microcavities which have a small
volume and thus a high quality factor. Suitable candidates for the
present proposal are, for example, the microtoroidal cavities which
has cooperativity factor $g^2/(\Gamma_C\Gamma_E)\sim10^7$ and the
ratio $g/\Gamma_E\sim10^3$ [23,24], where $g$ is defined by
$g=\mathrm{max}(g_a,g_b)$. Thus our scheme is feasible with current
available systems.

\section{Conclusion}

In conclusion, we have presented a scheme for nondestructive
identifying unknown BDS by measuring the probe qubits. This scheme
is implemented in three steps. In each step we can build an equation
between the unknown coefficients of the BDS and the observable of
the probe qubit. Combining the three equations we can calculate the
parameters. Moreover, at the end of each step, the BDS ensemble
remains in the initial state, therefore it is not polluted by the
identification processing. We also consider the experimental
realization of the scheme in the cavity QED system. By selecting
appropriate Rabi frequencies of the driving lasers, we can realize
the corresponding unitary operations, respectively. Our scheme is
feasible with the current techniques.

\section{Acknowledgement}

This work was supported by the National Natural Science Foundation
of China, under Grants No. 10805007 and No. 10875020, and the
Doctoral Startup Foundation of Liaoning Province.

\end{document}